\documentclass[12pt]{iopart}
\usepackage{iopams}
\usepackage{amssymb}
\usepackage{graphicx}
\usepackage{dcolumn}
\usepackage{bm}
\newcommand{\layx}[1]{^{[#1]}}

\begin{document}

\title[Determinants of public cooperation in multiplex networks]{Determinants of public cooperation in multiplex networks}

\author{Federico Battiston,$^{1,*}$ Matja{\v z} Perc,$^{2,3}$ Vito Latora$^{1,4}$}
\address{$^1$School of Mathematical Sciences, Queen Mary University of London, London E1 4NS, United Kingdom\\
$^2$Faculty of Natural Sciences and Mathematics, University of Maribor, Koro{\v s}ka cesta 160, SI-2000 Maribor, Slovenia\\
$^3$CAMTP -- Center for Applied Mathematics and Theoretical Physics, University of Maribor, Mladinska 3, SI-2000 Maribor, Slovenia\\
$^3$Dipartimento di Fisica ed Astronomia, Universit{\`a} di Catania and INFN, I-95123 Catania, Italy}
\ead{f.battiston@qmul.ac.uk}

\begin{abstract}
Synergies between evolutionary game theory and statistical physics have significantly improved our understanding of public cooperation in structured populations. Multiplex networks, in particular, provide the theoretical framework within network science that allows us to mathematically describe the rich structure of interactions characterizing human societies. While research has shown that multiplex networks may enhance the resilience of cooperation, the interplay between the overlap in the structure of the layers and the control parameters of the corresponding games has not yet been investigated. With this aim, we consider here the public goods game on a multiplex network, and we unveil the role of the number of layers and the overlap of links, as well as the impact of different synergy factors in different layers, on the onset of cooperation. We show that enhanced public cooperation emerges only when a significant edge overlap is combined with at least one layer being able to sustain some cooperation by means of a sufficiently high synergy factor. In the absence of either of these conditions, the evolution of cooperation in multiplex networks is determined by the bounds of traditional network reciprocity with no enhanced resilience. These results caution against overly optimistic predictions that the presence of multiple social domains may in itself promote cooperation, and they help us better understand the complexity behind prosocial behavior in layered social systems.
\end{abstract}

\maketitle

\section{Introduction}
Human cooperation is an evergreen puzzle \cite{nowak_11}, at the heart of which is the divide between the Darwinian desire to maximize personal benefits and our social instincts that dictate prosocial behavior. The later are particularly strong in humans, because without their evolution we would have had serious challenges in rearing offspring that survived \cite{hrdy_11}, and as a results would have likely died out as a species. Instead, we have acquired remarkable other-regarding abilities that have propelled us to dominance over all the other animals, to the point where today the biggest threat to us is ourselves.

The theoretical framework used most frequently to study cooperation among selfish individuals is evolutionary game theory \cite{sigmund_93, weibull_95, hofbauer_98, nowak_06, sigmund_10}, where the concept of a social dilemma captures the essence of the problem. In short, cooperation is costly, and it therefore weighs heavily on individual wellbeing and prosperity. One is thus torn between doing what is best for the society, and doing what is best for oneself. The public goods game is particularly apt in describing the dilemma \cite{perc_jrsi13, perc_pla16}. The game is played in groups, where individuals can decide between cooperation and defection. Those that decide to cooperate contribute an amount to the common pool, while defectors contribute nothing. All the contributions are multiplied by a synergy factor that takes into account the added value of a group effort, and the resulting public goods are divided equally among all group members irrespective of their strategy. Clearly the best individual strategy is defection. But if everybody decides to defect there will be no public goods. In order to avoid the tragedy of the commons in a society cooperation is thus needed \cite{hardin_g_s68}.

While the evolution of cooperation has been studied at great lengths in biology and sociology \cite{nowak_s06, rand_tcs13}, the problem became attractive for physicists after the discovery of network reciprocity \cite{nowak_n92b}, which manifests as the formation of resilient cooperative clusters in a structured population \cite{vainstein_pre01, abramson_pre01, kim_bj_pre02, holme_pre03, masuda_pla03, zimmermann_pre04, santos_prl05, santos_pnas06, fu_pla07, gomez-gardenes_prl07, masuda_prsb07, szolnoki_epl08, floria_pre09, fu_pre09b, lee_s_prl11, fu_prsb11, fu_jsp13, pereda_g17}. Cooperators in the interior of such clusters can survive at conditions that do not sustain cooperation in well-mixed populations. In fact, methods of statistical physics have recently been applied to subjects that, in the traditional sense, could be considered as out of scope. Statistical physics of social dynamics \cite{castellano_rmp09}, of evolutionary games in structured populations \cite{szabo_pr07, perc_bs10, pacheco_plrev14, wang_z_epjb15}, of crime \cite{orsogna_plr15, helbing_jsp15}, and of epidemic processes and vaccination \cite{pastor_rmp15, wang_z_pr16}, are only some examples of this exciting development. An important enabler for this has been the coming of age of network science \cite{barabasi_16}, which has been going from strength to strength during the past decade and a half \cite{albert_rmp02, newman_siamr03, boccaletti_pr06, fortunato_pr10, holme_sr12, kivela_jcn14, boccaletti_pr14, barabasi_16}, providing key theoretical foundations for modeling social systems.

We are here concerned with the evolution of cooperation in multiplex networks, which have, together with the closely related multilayer and interdependent networks, recently emerged as the new frontier in network science \cite{de2013mathematical, gomez2013diffusion, radicchi2013abrupt, bianconi2013statistical, nicosia2013growing, battiston14, hackett2016bond, battiston17, kivela_jcn14, boccaletti_pr14}. Indeed, multiplex networks are able to account for the variety of different social contexts an individual may be involved in, and are thus crucial for an in-depth understanding of human cooperation across different interaction layers \cite{wang_z_epjb15}. Several mechanisms have already been discovered by means of which the interdependence between different networks or network layers may help to increase the resilience of cooperation and resolve social dilemmas \cite{wang_z_epl12, gomez-gardenes_srep12, gomez-gardenes_pre12, wang_b_jsm12, wang_z_srep13, jiang_ll_srep13, szolnoki_njp13, wang_z_njp14, wang_pre14x, xia2015heterogeneous, meng2015spatial, luo2016cooperation, allen2017asynchronous}. Interdependent network reciprocity is one example, which requires simultaneous formation of correlated cooperative clusters on two or more networks \cite{wang_z_srep13}. Other mechanisms that promote cooperation beyond the bounds of traditional network reciprocity include non-trivial organization of cooperators across the network layers \cite{gomez-gardenes_srep12}, probabilistic interconnectedness \cite{wang_b_jsm12}, information transmission between different networks \cite{szolnoki_njp13}, as well as self-organization towards optimally interdependent networks by means of coevolution \cite{wang_z_njp14}.

Previous research has thus shown that multiplex networks may enhance the resilience of cooperation, but the key determinants of this, especially in terms of the topological overlap between the network layers and the game parametrization on each individual layer, still need to be determined. By studying the public goods game in a multiplex of regular random graphs, we here show that enhanced public cooperation requires significant edge overlap, combined with at least one layer being able to sustain some cooperation by means of a sufficiently high synergy factor. The details of this conclusion depend further on the number of layers forming the multiplex, and on other properties of the spatiotemporal evolutionary dynamics, which includes pattern formation and spontaneous symmetry breaking across the layers. As we will show, these results provide a deeper understanding of the complexity behind cooperation in multiplex networks, and as such they have important implications for promoting prosocial behavior in different but linked social contexts.

The organization of this paper is as follows. We present the definition of the public goods game in the multiplex and the details of the Monte Carlo simulation procedure in Section~II. Main results are presented in Section~III. We conclude with the summary of the results and a discussion of their implications in Section~IV.

\section{Public goods game in the multiplex}

In the public goods game players, belonging to a group of size $G$, are asked to contribute to a common pool.
Cooperators contribute with a token $d$, typically $d=1$, whereas defectors do not contribute at all. The amount of tokens in the pool is multiplied by a synergy factor $r$, and the resulting amount is divided equally among all players. Cooperators thus obtain a payoff $d ( N_C \cdot r/G -1)$, while defectors get
$d  N_C \cdot r/G$, where $N_C$ is the number of cooperators in the group.
When the game is played on multiple rounds, players choose to cooperate or defect at each iteration, based on the success of the two strategies.
The game can be implemented on structured populations, where players are placed on the nodes of a graph
and interact through their links. The results usually show that, while in well-mixed populations cooperators extinguish quickly,
repeated local interactions among the same players allow the formation of clusters of cooperators which are able to survive \cite{santos_n08, szolnoki_pre09c, gomez-gardenes_epl11, gomez-gardenes_c11, gracia2014intergroup, bottcher2016promotion}.

In real situations, individuals are typically involved in strategic
choices on independent domains, and can adopt different strategies
according to the specific domain.  However, information on the
earnings of each individual might be only be available at the
aggregate level, as a sum the payoff obtained as a result of all its
decisions.  More formally, in order to take this into account, we
consider here a population of $N$ individuals playing the public goods
game on the $M$ layers of a multiplex network. In particular, we model
each layer as a regular random graph with degree $k=4$. Hence the game
is played in groups each of size $G = k+1$. The state of a player $i$,
$i=1,2, \ldots, N$, is fully described by a vector of strategies ${\bm
  s}_i = \{ s_i \layx 1, \ldots, s_i\layx M \}$, such that at each layer
$\alpha$, $\alpha=1,2, \ldots, M$, the player can independently choose
to either cooperate, i.e., $s_i^{[\alpha]}=+1$, or defect, i.e.,
$s_i^{[\alpha]}=-1$.  The benefit of synergy among individuals in
general depends on the specific domain. In order to model this
feature, we assume that the synergy factor can be different from layer
to layer. We hence consider a synergy factor vector $\bm r = \{ r \layx
1, \ldots, r\layx M \}$. On each layer $\alpha$, player $i$ earns the
payoff $\pi_i^{[\alpha]}$, such that $i$ gains $d ( N_C \cdot
r\layx{\alpha}/G -1)$ if it cooperates, or otherwise $d N_C \cdot
r\layx{\alpha}/G$, if it defects.

The public goods game is simulated by a Monte Carlo method (for a fast implementation using parallel computing see Ref.~\cite{perc_ejp17}) in which,
at each elementary step, a layer $\alpha$ is selected, and then a
randomly chosen node $i$, and one of its neighbors $j$ on that layer,
are considered. Both $i$ and $j$ play the game on all the
layers, thereby obtaining respectively payoffs
$\pi_{i}=\sum_{\alpha=1}^M \pi_{i}^{[\alpha]}$ and
$\pi_{j}=\sum_{\alpha=1}^M \pi_{j}^{[\alpha]}$.  Finally, player $i$
compares its payoff to that of player $j$, and
copies the strategy of player $i$, but only at the layer $\alpha$,
with a probability given by a Fermi function:
\begin{equation}
W(s_i^{[\alpha]} \to s_j^{[\alpha]} )= \biggl (1+ exp \biggl
[\frac{\pi_j - \pi_i}{K} \biggr ] \biggr)^{-1},
\end{equation}
where $K$ quantifies the contribution of random fluctuations to the
strategy adoption \cite{szolnoki_pre09c, javarone16}. In
the $K \to 0$ limit, player $j$ copies the strategy of player $i$ if
and only if $\pi_i > \pi_j$. Conversely, in the $K \to \infty$ limit,
payoff differences cease to matter and $i$ copies the strategy of $j$
with a probability equal to 0.5. Between these two extreme cases, for
intermediate values of $K$, players with a higher payoff will be
readily imitated, although the strategy of under-performing players
may also be occasionally adopted to mimic, for example, errors in the
decision making, imperfect information and external influences that
may adversely affect the evaluation of an opponent. We adopt the value
$K=0.5$ without loss of generality, as shown in \cite{szolnoki_pre09c}.
In our simulations, we obtain one full Monte Carlo step (MCS) by
repeating $M \times N$ times the elementary steps described above,
thus giving a chance to every player to change its strategy on all the
layers once on average.

In order to characterize the outcomes of our evolutionary
dynamics model, we
introduce the vector $\bm c = \{ c \layx 1, \ldots, c \layx M \}$, where
$c \layx \alpha$ is the fraction of cooperators at layer $\alpha$ in
the stationary state, i.e., when the average over time of this
quantity becomes time independent. As a first order parameter we consider
then the quantity
\begin{equation}
c = \frac{1}{M}\sum_{\alpha=1}^M c \layx \alpha,
\end{equation}
which is the overall fraction of cooperators across all the layers of the
multiplex in the stationary state. We note that $0 \le c \le 1$, where $c=1$
corresponds to full cooperation while $c=0$ corresponds to full
defection. As a second order parameter, we define the average
coherence $\xi$ of the players across all the layers,
defined as:
\begin{equation}
\xi=\frac{1}{N}\sum_{i=1}^N \xi_i  \qquad \xi_i= \left|\frac{1}{M}\sum_{\alpha=1}^M s_i\layx \alpha \right|
\end{equation}
where $\xi_i$ is the coherence of the strategies of player $i$.
A value $\xi_i=1$ indicates that player $i$ is maximally coherent,
meaning it adopts the same strategy in all the layers. Conversely,
$\xi_i=0$ means that player $i$ is maximally incoherent, adopting
$s_i^{[\alpha]}=+1$ just as often as $s_i^{[\alpha]}=-1$ across the
$M$ different layers. A similar definition of coherence for the
particular case $M=2$ has been reported in Ref.~\cite{battiston16_interplay}.
In addition to the two main order parameters $c$ and $\xi$, when $M=2$, we will
also use the quantities $c \layx 2 - c \layx 1$ and $|c \layx 2 - c \layx 1|$
to evaluate differences in the level of cooperation at the two layers. In fact,
as we will show in the following, there exist indeed regions in the
parameter space of our model such that full cooperation is observed at
the layer with the highest synergy factor, while
the layer with the lower synergy factor is
in a full defection state. Also, a multiplex network with two layers
may exhibit spontaneous symmetry breaking, such that, even if the two layers are
characterized by the same synergy factor and the same interaction network topology,
the level of cooperation on them can be different.

\section{Results}

\begin{figure}
\center
\includegraphics[width=12cm]{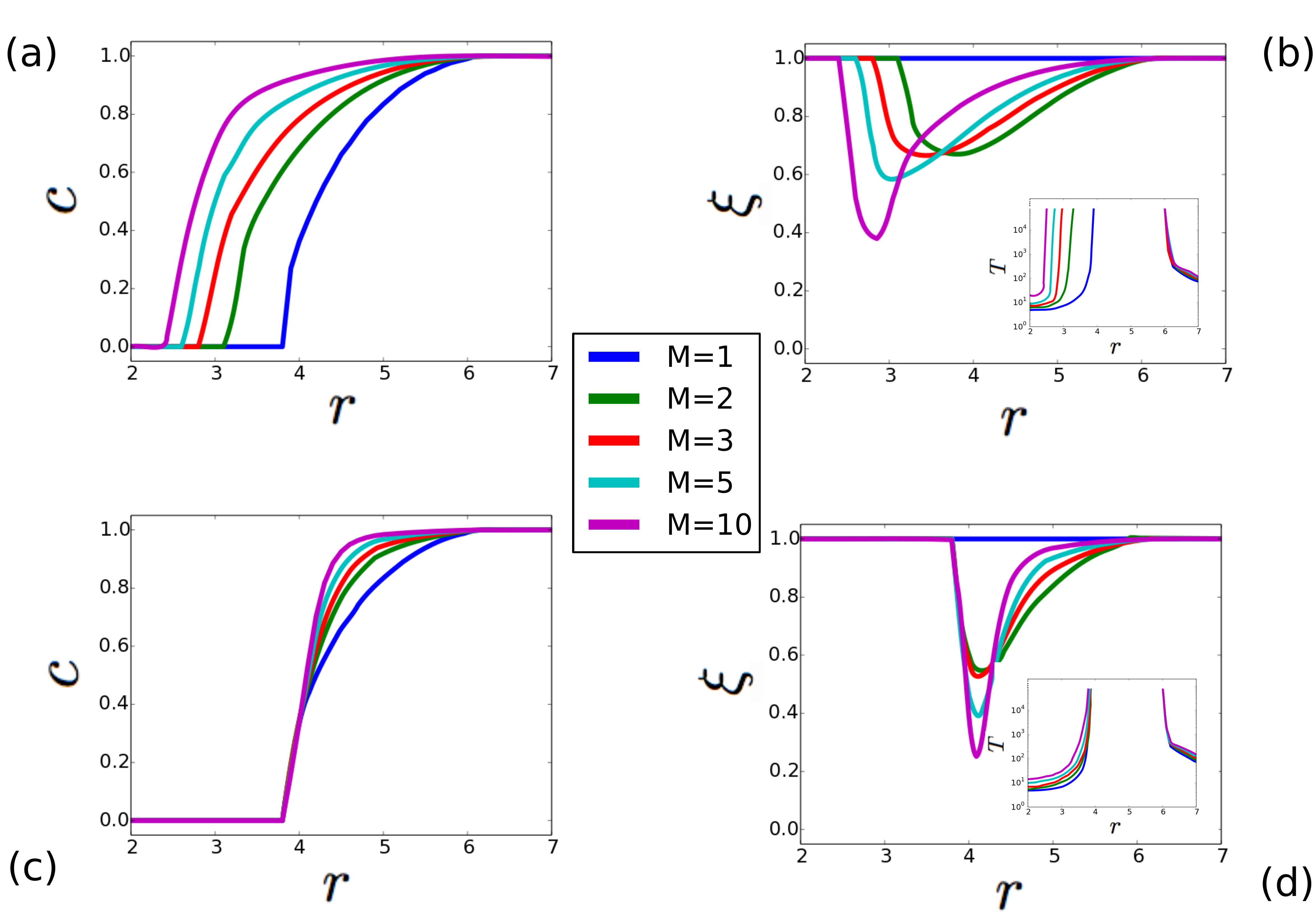}
\caption{Number of layers and topological overlap are crucial to lower
  the critical value of the synergy factor needed
  for cooperators to survive. (a,b) The multiplex is formed by $M$
  regular random graphs with degree $k=4$, and edge overlap respectively
  equal to $\omega=1$ (a,b) and $\omega=0$ (c,d).
  We show the average fraction of cooperators
  across the whole multiplex $c$ (a,c) and the average coherence of
  the players across all the layers $\xi$ (b,d) as a function of the
  synergy factor $r$, and for different values of $M$. Insets
  show the number of full Monte Carlo steps $T$ needed for the system
  to reach an absorbing phase with either all cooperators or all defectors.}
\label{multilay}
\end{figure}

We implement our model on a multiplex network in which it is possible
to tune the similarity among the topology of the $M$ layers. We therefore consider
that each layer is a regular random graph with $N = 10^4$ nodes and
$K=2 \cdot 10^4$ links, and we tune the
average edge overlap $\omega$
of the multiplex~\cite{bianconi2013statistical, battiston14}. Such overlap is defined as the average of all the edge overlaps computed between pairs of layers $\omega=\frac{2}{M(M-1)}\sum_{\alpha, \beta > \alpha}^M \omega^{[\alpha,\beta]}$, where:
\begin{equation}
\omega^{[\alpha,\beta]} =\frac{\sum_{i,j>i} a_{ij}\layx{\alpha} a_{ij}\layx{\beta} }{  \sum_{i,j>i} (a_{ij}\layx{\alpha}  + a_{ij}\layx{\beta} - a_{ij}\layx{\alpha} a_{ij}\layx{\beta}   ) }.
\end{equation}
When all layers are equal the topological overlap is maximum and $\omega=1$. Conversely, if there are no
pairs which are connected on more than one layer, the overlap is
minimum and $\omega=0$.
We begin by assuming that the synergy factor used for the
public goods game is the same at each layer, namely we set
$r\layx \alpha = r \quad \forall \alpha$. Initially, each layer is
populated by the same proportion of
cooperators and defectors, distributed uniformly at random, and subsequently
the game is iterated in time according to the Monte Carlo simulation
procedure described in Section II.  Results presented in
Fig.~\ref{multilay} are shown separately in two rows, respectively for the case
$\omega=1$ [panels (a) and (b)] and the case $\omega=0$ [panels (c) and (d)].
Looking at panels (a) and (b), it can be observed that the larger the value of
$M$, the lower the critical value of the synergy factor that is needed
to sustain cooperation. When $M=1$, on a single-layer regular random
graph, the critical value is equal to $r_c=3.75$, which is in
agreement with traditional network reciprocity
\cite{szolnoki_pre09c}. When ten layers form the multiplex, however,
the critical value drops to as low as $r_c=2.35$. The minimal
coherence also emerges at ever lower values of $r$ as $M$ increases,
and the minima become lower, indicating that at least some layers are
able to sustain cooperation even though in the majority the players
defect.

The evolutionary outcomes are significantly different in panels (c)
and (d), where the topological overlap is zero. It can be observed
that the increase in $M$ does nothing to reduce the critical values of
$r$ needed to sustain cooperation. In fact, $r_c=3.75$ that is due to
traditional single-layer network reciprocity always emerges as the
necessary condition for public cooperation. Not surprisingly, the
minimal coherence also occurs at the same value or $r$ regardless of
$M$. The minima become lower as $M$ increases because the average goes
over more layers, simply giving statistically more opportunity for
players to hold different strategies across different layers. Taken
together, these results show that topological overlap is essential for
enhanced multiplex network reciprocity to take effect and enhance the
resilience of public cooperation expected in a system with multiple layers of interactions. These results confirm the crucial
impact of the edge overlap on dynamical processes on networks, in
agreement with previous works~\cite{battiston16_axelrod, battiston16_exploration}.

\begin{figure}
\center
\includegraphics[width=14cm]{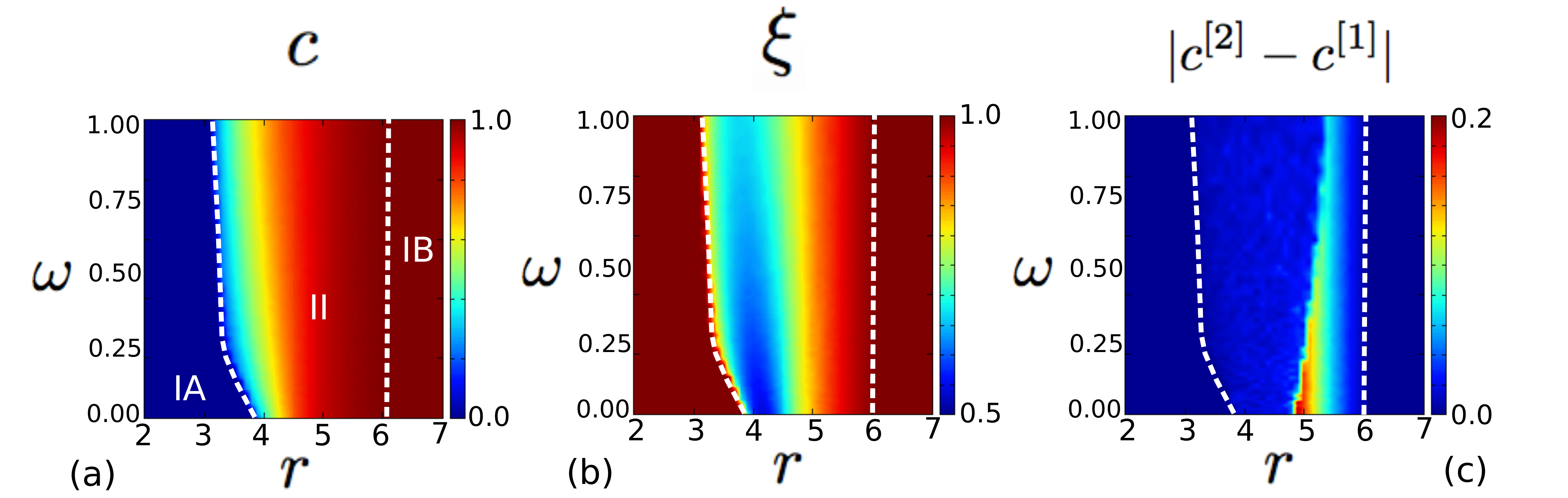}
\caption{Benefits to public cooperation in a multiplex network with
  $M=2$ layers and a tunable value of edge overlap $\omega$.  The
  synergy factor used for the public goods game is the same at the two
  layers, namely $r \layx 1 = r \layx 2 = r$. Panel (a) shows the full
  $r-\omega$ phase diagram where the color map encodes the average
  fraction of cooperators $c$ across the two layers. High benefits
  emerge only for large values of $\omega$.
  Panels (b) and (c) show the full
  $r-\omega$ phase diagram where the color map encodes respectively the average
  coherence  $\xi$ and the absolute difference $|c \layx 2 - c \layx 1|$
  between the fraction of cooperators in the two layers. Reported dashed
  lines separate regions where the system is at an absorbing state with
  full coherence $\xi=1$ and either full defection $c=0$ (IA) or
  full cooperation $c=1$ (IB), from the region of continuously evolving
  coexistence of cooperators and defectors $0<c<1$ (II).
}
\label{overlap}
\end{figure}

To investigate more in details the role of the topological overlap, we
consider a multiplex with only two layers, but where the value of the edge overlap
$\omega$ can be varied continuously in the range $[0,1]$.
In order to tune $\omega$ we have start with a configuration with  $\omega=1$,
where the two layers are made by the same regular random graphs with $k=4$.
Keeping fixed the structure of the first layer, we then start rewiring a fraction $f$
of the links on the second layer, so that we result in a
network with an edge overlap equal to:
\begin{equation}
\omega = \frac{(1- f)}{(1+f)}
\end{equation}
as a function of $f$
(see Refs.~\cite{diakonova16, battiston17} for details).
In particular, when $f=1$ and all the links of the second layer are rewired we get
with a multiplex network with no edge overlap, i.e. with $\omega=0$.
In Fig.~\ref{overlap} we report the results obtained as a function of the two
control parameters $r$ and $\omega$.
The three phase diagrams shown encode respectively the average fraction of
cooperators $c$ across the two-layer multiplex (a), the average
coherence $\xi$ of the players (b), and the
absolute difference between the fraction of cooperators in the two
layers, $|c \layx 2 - c \layx 1|$ (c). In panels (a) and (b), we can
distinguish two regions, namely type I where the whole multiplex
reaches an absorbing phase (each layer is either in full cooperation
or in full defection), and type II where the multiplex is trapped in a
state where cooperators and defectors coexist. The two type I regions
can be further classified as type IA where defectors dominate ($c=c
\layx 1 = c\layx 2 = 0$), and type IB where cooperators dominate ($c=c
\layx 1 = c\layx 2 = 1$). In both type IA and IB regions all the players
are of course fully coherent, i.e., they adopt the same strategy on
both layers such that $\xi=1$. It can be observed that the added value of the multiplex structure in enhancing network reciprocity marking the transitions from region IA to II suddenly decreases for $\omega<0.25$, disappearing as the value of the topological overlap $\omega$ approaches 0.

Interestingly, at the II to IB transition, that is from the mixed $(C+D)$ phase to the pure $C$ phase, the topological overlap does not play a role at all, indicating that the enhanced multiplex network reciprocity is crucial only when cooperation can be barely sustained. Even as the multiplex enters the mixed $(C+D)$ phase, i.e., region II, the impact of the extent of topological overlap vanishes very quickly beyond the critical value of $r_c$ at the transition point. In the mixed $(C+D)$ phase, we can also observe spontaneous symmetry breaking in panel (c), where in region II $|c \layx 2 - c \layx 1|>0$. This means that, even though the public goods game in both layers is characterized by the same synergy factor and is staged on layers with identical topological properties, the level of cooperation in the stationary state is different. In particular, it can be observed that the lower the topological overlap between the two layers ($\omega \to 0$), the higher the symmetry breaking, with the maximum value occurring for $\omega=0$ and $r=G=5$.

\begin{figure}
\center
\includegraphics[width=14cm]{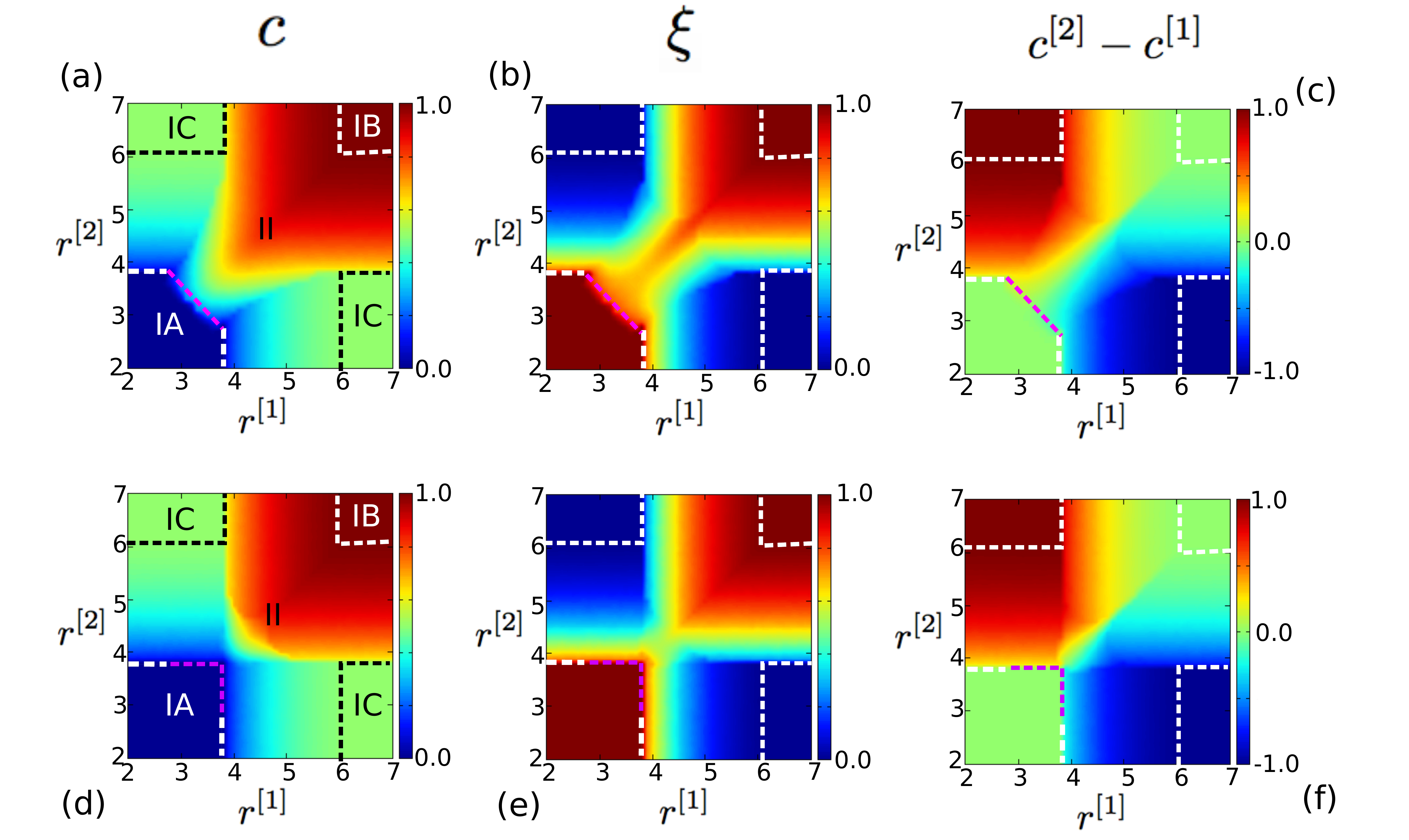}
\caption{The emergence of multiplex network reciprocity depends on the
  different values of the synergy factors at each layer.  The
  multiplex is formed by two layers of regular random graphs with an
  edge overlap respectively equal to $\omega=1$ (a,b,c) and $\omega=0$
  (d,e,f).  Panels (a,d) report the full $r\layx 1 - r\layx 2$ phase
  diagram where $r\layx 1$ and $r\layx 2$ are the synergy factors at the
  two layers, and where the color map encodes the average fraction of
  cooperators $c$. Panels (b,e) and (c,f) show the full $r\layx 1 -
  r\layx 2$ phase diagram where the color map encodes respectively the
  average coherence $\xi$ and the difference of cooperators $c \layx 2
  - c \layx 1$ in the two layers. A new absorbing state (region IC),
  with full cooperation at layer $\alpha$ $c\layx \alpha = 1$, full
  defection at layer $\beta$ $c\layx\beta=0$ and complete incoherence
  $\xi = 0$, emerges.  }
\label{diffr}
\end{figure}

Lastly, we study the impact of differing synergy factors in the layers
forming the multiplex in order to determine the importance of game
parametrization on the emergence of enhanced multiplex network
reciprocity. In Fig.~\ref{diffr}, we present results separately for
two-layer multiplex networks with complete (a,b,c) and zero (d,e,f)
topological overlap between the layers. The $r\layx 1 - r\layx 2$ phase
diagrams encode the average fraction of cooperators across the
two-layer multiplex $c$ (a), the average coherence of the players
across the two layers $\xi$ (b), and the difference between the
fraction of cooperators in the two layers $c \layx 2 - c \layx 1$
(c). It can be observed that in both cases, regardless of the overlap,
region IB occurs when both $r\layx 1 > 6$ and $r\layx 1 > 6$ (a,d). A
new region can also be observed in both cases when $r\layx \alpha > 6$
and $r\layx \beta < 3.75$, which we denote as IC, where one layer is
characterized by full cooperation ($c\layx \alpha=1$), while the other
layer is characterized by full defection
($c \layx \beta=0$). Accordingly, we have a completely
incoherent multiplex with $\xi=0$ (b,e).

These equivalences beget the question when do the evolutionary outcomes actually differ in dependence on complete and zero overlap. As results in Fig.~\ref{diffr} show, and as could be anticipated from the results presented in Fig.~\ref{overlap}, the difference is most expressed at the interface between regions IA and II (see dashed purple line). If there is no topological overlap between the two layers (d,e,f), we see that as long as both $r\layx 1 < 3.75$ and $r\layx 2 < 3.75$, we always have full defection in both layers. Hence, multiplexity does not provide any advantage to the evolution of cooperation (see also Fig.~\ref{multilay}). Conversely, when there is perfect overlap (a,b,c), cooperators emerges already for $r = r\layx 1 = r\layx 2 < 3.75$, roughly $r \approx  3.25 = r_c$. But given an arbitrary choice for $r\layx \beta$ that is smaller than $3.75$, is $r\layx \alpha=3.25$ the minimum value to see the emergence of cooperators in the multiplex? Indeed no, given $r\layx \beta$ we still see cooperators in the system as long as we choose $r\layx \alpha$ such that it is slightly above $\frac{r\layx \alpha + r\layx \beta}{2} > 3.25$ (linear relationship $r_c\layx \alpha = 2 \cdot 3.25 - r\layx \beta$, see purple line). Importantly, this relation holds as long as $r\layx \beta$ does not go beyond $3.75$, the original critical value for one layer, at which point the relation no longer holds and the overlapping case behaves as the non-overlapping case. Based on the phase diagrams in Fig.~\ref{diffr}, the critical value can be approximated as $r\layx \alpha \approx 2 \cdot r_c - r_c(M=1)$, which means $r\layx \alpha = 2 \cdot 3.25 - 3.75 \approx 2.75$. Taken together, a topologically overlapping multiplex can extend the coexistence region II towards significantly smaller values of $r$, which in our two-layer setup corresponds to a triangle delimited by $(r\layx 1, r\layx 2)$ such that $X=(2.75,3.75)$, $Y=(3.75,3.75)$ and $Z=(3.75,2.75)$.

\section{Discussion}
We have studied the determinants of public cooperation in multiplex networks, focusing in particular on the topological overlap and different synergy factors across the layers. We have shown that, if the topological overlap between the layers is sufficiently extensive, the critical value of the synergy factor that enable cooperators to survive decreases steadily as the number of layers increases. This result confirms the existence of interdependent or multiplex network reciprocity, which enhance the resilience of cooperators beyond the bounds of traditional network reciprocity on an single-layer network. However, we have also shown that, as the topological overlap between the layers decreases, so do the benefits of multiplexity for the evolution of cooperation. In particular, if the topological overlap is zero, cooperators loose all benefits stemming from their engagement in different layers of the multiplex and thus become reliant on single-layer network reciprocity alone. These results manifest not only in the average fraction of cooperators in the multiplex, but also in the average coherence of the players across all the layers. We show that, in case of perfect topological overlap, the later reaches a minimum at ever lower values of the synergy factor as the number of layers increases, while in the absence of topological overlap the synergy factor yielding minimal coherence is independent of the number of layers.

By further varying the synergy factor that applies on each particular layer, we have shown that the topological overlap is crucial only if the synergy factor on all layers is smaller than the critical value on a single layer. If that is the case, the overlap plays a key role in sustaining cooperation, and there exists an average value of the synergy factor across all the layers that needs to be reached for cooperators to survive. However, if on a single layer the synergy factor is large enough to sustain cooperation even in the absence of multiplexity, i.e., as if the layer would be isolated, then the topological overlap seizes to matter. By means of extensive Monte Carlo simulations, we have determined precise bounds on the topological overlap and the relations between synergy factors in different layers that need to be met for enhanced multiplex network reciprocity to take effect. Taken together, our results thus establish key determinants of public cooperation in multiplex networks.

The presented results reveal rather stringent conditions that have to be met for public cooperation to be more resilient on multiplex network than it is on single-layer networks. Indeed, the hallmark of multiplex network reciprocity, which is the simultaneous formation of correlated cooperator clusters across different layers, requires near perfect topological overlap, and is effective only if the conditions for cooperation on all layers are rather dire. If the coordination process leading to the forlimation of correlated cooperator clusters is disturbed due to the lack of topological overlap, multiplex network reciprocity never emerges, resulting in the total collapse of cooperation across all layers minus those that would sustain cooperation on their own either way. Thus, while multiplexity and network interdependence can in theory be exploited effectively to promote cooperation past the limits imposed by isolated networks, caution is needed against overly optimistic predictions that suggest involvement in different social contexts alone is in itself sufficient to promote cooperation. Enhanced prosocial behavior in layered social systems can emerge only if the positions of the players and the links among them in these layers do not differ much.

\ack
V.L. acknowledges support from EPSRC projects EP/N013492/1. M.P. is supported by the Slovenian Research Agency (Grants J1-7009 and P5-0027).

\section*{References}
\providecommand{\newblock}{}

\end{document}